# Computer-assisted workflows composition based on Virtual Simulation Objects technology

Pavel A. Smirnov, Sergey V. Kovalchuk, Alexander V.Boukhanovsky
*E-Science Research Institute*
*ITMO University, St.Petersburg, Russia*
smirnp@gmail.com, sergey.v.kovalchuk@gmail.com

**Abstract**

The existing approaches for scientific workflows composition face the problems of domain knowledge integration. By this paper we summarize the results, which have been elaborated and implemented during the 2-year research concerning to Virtual Simulation Objects (VSO) concept and technology development. The contribution of this paper consists of formal models of the VSO internal structures and user-assistance logic, which may be obtained as a result of the reasoning over knowledge base.

**Keywords:** scientific experiment, workflow composite application, domain knowledge integration

## 1. Introduction

The existing E-Science paradigms' [1] propose the idea, that virtually simulated experiments may deliver novel scientific results' because such experiments could not be reproduced in real life (environment hazards, epidemics dispersion, meteorological forecasting & etc. are the examples). In order to solve these complex tasks more effectively domain experts are often required to collaborate with colleagues from adjacent domains, so the experiments became multidisciplinary. The fourth paradigm [2] of E-Science proposes the collaboration of geographically distributed groups of scientists (called virtual organizations) in order to solve multidisciplinary problems. Workflow-based computational experiment is the most spread way to organize distributed investigations. But one the main difficulties, which scientists face is necessity to have an expertise for design new workflow-applications or reuse already existing ones. These processes require the user to be aware not only about domain knowledge (methodology and algorithms), but also about the problem knowledge (dataflow language, packages and resources of an execution platform). As a result, workflow-based applications are platform-dependent and are not easy to be reproduced or reused by other scientists. The solution of this problem requires some mechanisms for knowledge formalization, sharing and publication in order to make domain-specific tools available for multidisciplinary investigations. Some of existing attempts performed in this field are presented in the next section. In our previous paper [3] we have proposed Virtual Simulation Objects (VSO) concept and technology as a solution of the problem. Motivation of this paper is to summarize the results, which have been elaborated during the 2-year research. The contribution of this paper is a formal model of VSO-structures and formal description of user-support functionalities, provided due to reasoning over knowledge base.

## 2. Related work

The idea of user-assistance support for composite applications design is not new. There are several projects, which apply knowledge-based technologies at different stages workflow-application's lifecycles: design, storage, provenance analysis. The following papers propose an idea of workflow composition via abstract workflow-candidates [4], conceptual fragments [5], model of computations [6] and functional units [7].

Paper [4] proposes an idea of abstract workflow which is formed by user-specified requests (called seeds) – sets of constraints, parameter configurations or dataset selections. The particular implementation (called executable workflow) is formed after automatic search and encapsulation of executable workflow templates into seed's structures. The WINGS system searches workflow-candidates over template-catalog and verify them according to specified constraints. The WINGS assumes access of external software catalogs and datasets through service endpoints.

Paper [5] describes composition on conceptual (meta-) workflow level and its implementation, organized through abstract fragments. Fragments consist of a pair of pattern and blueprint. The blueprint is an operation (or set of operations), which should be applied within pattern's structure. Executable workflows are generated according to model-driven knowledge, formalized during knowledge-capturing mapping process. The generation is implemented in semi-automatic mode via advanced pattern matching techniques.

In contrast to the ideas of these two papers, a VSO-concept do not deals with templates or fragments (candidates or blueprints), which should be matched to the specified requirements in order to fill the some abstract workflow structures (seeds or patterns). The idea of VSO proposes model-driven knowledge-based design process using high-level configurable virtual objects with formalized low-level knowledge. Virtual objects are specially structured semantic models and consequence of these models allow to perform some comparison operations over each other. Due to semantic-web technologies implementation a VSO technology provides user-assistance during knowledge formalization and environment composition processes. Automatic connection of semantically equivalent parameters performs the assistance logic during composition process.

Paper [6] presents the models of computation defined by the selected director. Director is an entity, which defines the semantics of behavior between two actors, connected by the director. Sets of directors and their semantics are defined by workflow-designer. This idea seems more similar to VSO concept, than the two previous ones, because it defines a format of collaboration between the executable components. VSO similarly offers a set of predefined methods consisting of executable packages. But VSO models are wider and propose composition at the top level of abstract, hiding the executable packages level from user.

Also the similar model-driven approach is presented in paper [6], where executable elements (software packages or services) are annotated as belonging to functional unit performing single or polymorphic or composite invocation pattern. Units should be defined manually and semi-automatic mode. Comparing this idea with VSO we can note, that constitution of functional consequence of low-level elements is performed in VSO, but only at the stage of method's definition. Methods directly are not used for composition, but the configurable high-level entities with the included computational models are used instead. Methods only perform an invocation patterns within complex VSO-structures.

The overview of contemporary situation in field of use-assisted workflow-composition techniques have shown that the idea to design workflow-based applications via composition of high-level entities with semantic-equivalence of their parameters' still have not been offered elsewhere. Construction of virtual system with a set of virtual objects looks rather perspective for automatic workflows generation, validation and reuse. Virtual objects are template-independent reusable building blocks, which couple the formalized domain and problem knowledge.

### 3. Knowledge usage processes

To demonstrate the logic of key processes required and provided by VSO-toolbox the following IDEF-0 process-flow diagram (Fig.1) was designed:

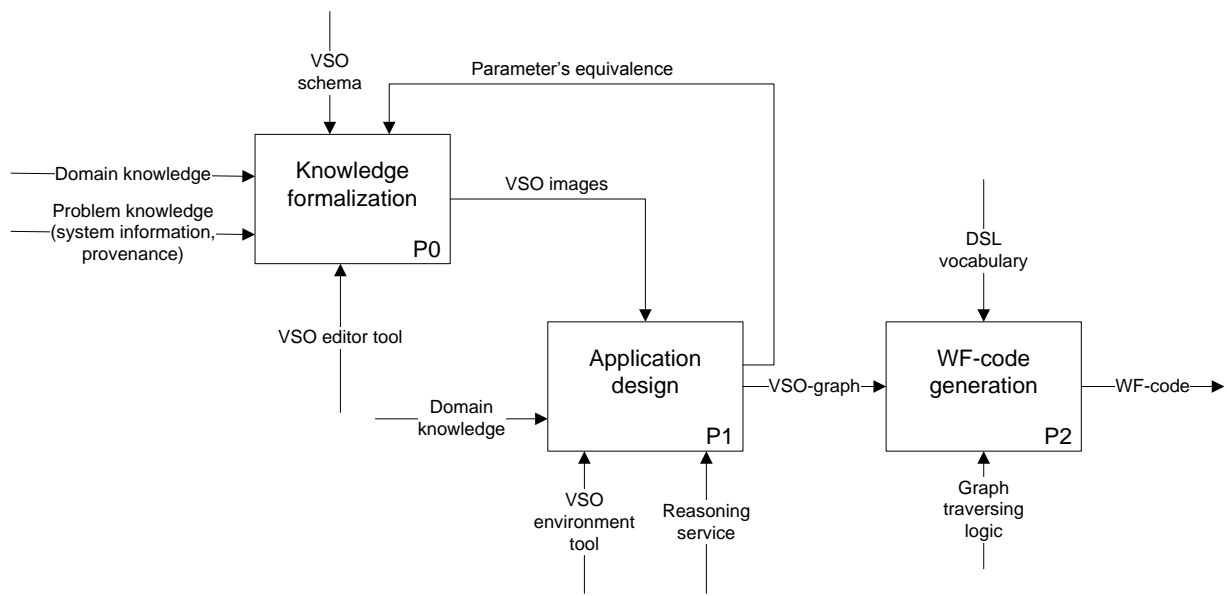

**Figure 1. Knowledge usage diagram**

Composite application design requires a set of predefined virtual object's images, which may be obtained as a result of knowledge formalization process (P0 at fig.1). This process is oriented for scientists, who have already implemented their software packages and are aware about workflow. The meaning of the P0 process is to make domain experts to specify the logic of their workflows into fine-grained VSO structures (the structure definition will be presented in the next section). These structures will be stored in knowledge base and used by other scientists during the next stage (P1 at fig.1). The consequence between packages and parameters of their values constitutes the domain knowledge. Definition of software packages, nodes where these packages are available, performance models and any platform-dependent service information constitute the so-called problem knowledge. The separation of domain and problem knowledge is also known as the Separation of Concerns and described in [**5**].

The second stage is virtual environment design process or composite application design (see P1 at fig.1). This process is oriented on scientists, who already do not have expertise about the underlying consequences of executable software packages. The only thing users should do – to drag and drop the VSO-images from catalog into environment. A separate instances of desired images will be automatically instantiated into virtual environment with the automatic connections between semantically equal input/output parameters. Application design process goes with reasoning service, which provides certain intelligent user-support functionality, described in the next section of the paper. The application design process may provide a certain novel domain knowledge concerning the user-defined consequences of objects', their parameter's equivalence which is also will be formalized and saved into knowledge base.

The final stage workflow generation process (see P2 at fig.1). Workflow script is generating from VSO-graph automatically with the application of graph traversing logic and domain-specific language vocabulary. VSO-graph consists of several object's instances interconnected via semantic-equivalent input/output parameters, which equivalence originally is defined by user.

## 4. The formal model

Thought the first paper [3] about virtual simulation objects and technology contains a methodological description, the detailed formal description of its entities haven't been presented yet. In this section the basic and extended formal models are presented:

**1) The basic formal model** describes a hierarchical structure of virtual object and consists of the following entities, mentioned in [3] :

*Simulated object* is the main entity to be operated during knowledge formalization and application design processes described before. Formally the virtual simulated object is described by the following tuple:

$$vso = \langle P, M, IN_{vso}, OUT_{vso} \rangle \qquad (1)$$

where $P = \{p\}$ is a set of object's properties, $M = \{m\}$ is a set of simulation models performing virtual representation of investigating object, $IN_{vso} = \{in_{vso}\}$ and $OUT_{vso} = \{out_{vso}\}$ are sets of input/output (IO) parameters respectively. These sets are defined via the conjunction of object's properties in the following way:

$$IN_{vso} = \bigcup_{i \to j} IN_{m_j} \cup P \qquad (2)$$

$$OUT_{vso} = \bigcup_{i \to j} OUT_{m_j} \cup P \qquad (3)$$

where $j$ is a capacity of $\{m\}$.

*Simulated model* describes a set of static and dynamic properties required for virtual object's simulation. The structure of single simulated model is defined as follows:

$$m = \langle S, IN_M, OUT_M \rangle \qquad (4)$$

where $S = \{s\}$ is set of methods available as implementation for model $m$, $IN_M = IN_{s'}$ and $OUT_M = OUT_{s'}$ are sets of IO parameters defined by corresponding sets of selected method $s'$.

*Method* is an imperative description of simulation algorithms implemented into software to solve the particular simulation task. Formal description of single method is presented by the following structure:

$$s = \langle IP, IN_s, OUT_s \rangle \quad (5)$$

where $IP = \{ip\}$ is a set of implementing packages, which constitute a consequence of execution packages for the method implementation, $IN_S = \{in_s\}$, $OUT_s = \{out_s\}$ are sets of input/output parameters defined as conjunction of corresponding parameters of implementing packages:

$$IN_s = \bigcup_{i \to k} IN_{ip_i} \quad (6)$$

$$OUT_s = \bigcup_{i \to k} OUT_{ip_i} \quad (7)$$

where $k$ is a capacity of $\{ip\}$.

*Implementing package* performs is platform-independent abstract package, which inherits the platform-dependent structure of really-executed package instantiated in distributed environment. The structure of single implementing package is described as follows:

$$ip = \langle sp, IN_{ip}, OUT_{ip} \rangle \quad (8)$$

where $sp$ is formal description of really-executed package instantiated in distributed environment, $IN_{ip} = \{in_{ip}\}$ and $OUT_{ip} = \{out_{ip}\}$ are sets of IO parameters defined by inheritance and extension of corresponding parameters of really-executed package $sp$ with default values and bindings to some semantic entity $Uri$:

$$in_{ip} = \langle in_{sp}, Value, Uri \rangle \quad (9)$$
$$out_{ip} = \langle out_{sp}, Uri \rangle \quad (10)$$

The two parameters are semantically equal, if they are bound with to same $Uri$, otherwise if the entities of their URIs are connected via "sameAs" property.

*Software package* is really-executable package within the distributed platform. In general case it may be defined as follows:

$$sp = \langle IN_{sp}, OUT_{sp} \rangle \quad (11)$$

where $IN_{sp} = \{in_{sp}\}$ and $OUT_{sp} = \{out_{sp}\}$ are sets of IO parameters. The formal description of really-executed package parameters is also platform-dependent and in general case may be defined as follows:

$$in_{sp} = \langle varname, value \rangle \quad (12)$$
$$out_{ip} = \langle varname \rangle \quad (13)$$

where $varname$ and $value$ are parameter name and value respectively.

**2) The extended formal model** broadens the basic one with composite objects' entities. A system of several composed objects may be presented as a single composite object and be composed with other composite objects, providing user with a macro-level simulation. Such functionality requires modification the formulas (1-3) with the following way:

$$vso = \langle P, VSO, M, IN_{vso}, OUT_{vso} \rangle \quad (14)$$

where $VSO = \{vso\}$ – is a set virtual objects included into composite object's structure. Objects inclusion is performed on the same logical level as model inclusion, what's why sets of IO parameters of composite objects will be extended with the following way:

$$IN_{vso} = \bigcup_{i \to j} IN_{M_j} \cup \bigcup_{i \to k} IN_{VSO_k} \cup P \qquad (15)$$

$$OUT_{vso} = \bigcup_{i \to j} OUT_{M_j} \cup \bigcup_{i \to k} IN_{VSO_k} \cup P \qquad (16)$$

where $j$ is capacity of $M$, $k$ is a capacity of $VSO$.

## 5. User-assisted functionalities

The formal description, presented below allows to describe the logic of intelligent user-support functionalities. Due to applying semantic-web technologies reasoning mechanism provides following knowledge-based user-support functionalities:

*1. Parameters' generalization* is one of the key functionalities required during application design process. Parameters generalization allows user to configure dataflows between objects and models within them. In general case the IO parameters of objects are recursively generalized from IO parameters of underlying implemented packages, which constitute the selected methods for simulated models implementations (see consequence between formulas 2-3, 6-7, 9-10). In the real use-case (see the references in the next section) at the upper "Objects" level (see fig.2) it's highly desirable to display the parameters, which values of bindings to other parameters have not been specified yet. Otherwise the application design process becomes highly overloaded with the amount of IO parameters, obtained as a result of generalization through 3 underlying levels. For example (see fig.2), all of implemented packages $IP_1 - IP_{10}$ will have at least one pair of IO parameters (one input and one output are the mandatory parameters for any package), the object $O_1$ at "Objects" level will have at 5-7 pairs of generalized IO parameters. To connect $O_1$ with $O_2$ at least one pair of IO parameter is required, so the rest 4-5 pairs are redundant to be shown at "Objects" level. The right solution here is to generalize only inputs (of all implementing packages in the consequence) with unspecified values and the outputs (of final package in the consequence only). So, by default all of the intermediate (specified or bound) IO parameters should be hidden at "Object level". This feature requires an some extension formulas 2, 6 and 3,7 with a following filtration condition respectively:

$$IN_A = IN_A \cap \{x \in IN_A, such\ that\ x.Value \in \emptyset\} \qquad (14)$$
$$OUT_A = OUT_A \cap \{x \in OUT_A, such\ that\ \exists Conn(x,T) = \emptyset\} \qquad (15)$$

where $A \in \{vso, m, s\}$ is a set of elements to be filtered, $Conn(x,T)$ is connectivity-relation with some other parameter T within the same abstraction level (methods, models, objects). The more details about connectivity-relations are presented in the next paragraph.

*2. Connections implication thought different abstraction levels* is the second significant feature provided by VSO concept and technology. The idea is that connections between elements at lower levels of abstraction recursively cause the connections between corresponding elements at the upper levels. Formally the logic of connection's implications occurred at the upper levels may be specified as follows:

$$Conn_{IP}\langle ip_1, ip_2 \rangle \Rightarrow Conn_S \langle s_1, s_2 \rangle \Rightarrow Conn_M \langle m_1, m_2 \rangle \Rightarrow Conn_{VSO} \langle vso_1, vso_2 \rangle \qquad (16)$$

To prove this statement we will demonstrate a single calculation cycle for some example structure from the figure 2 :

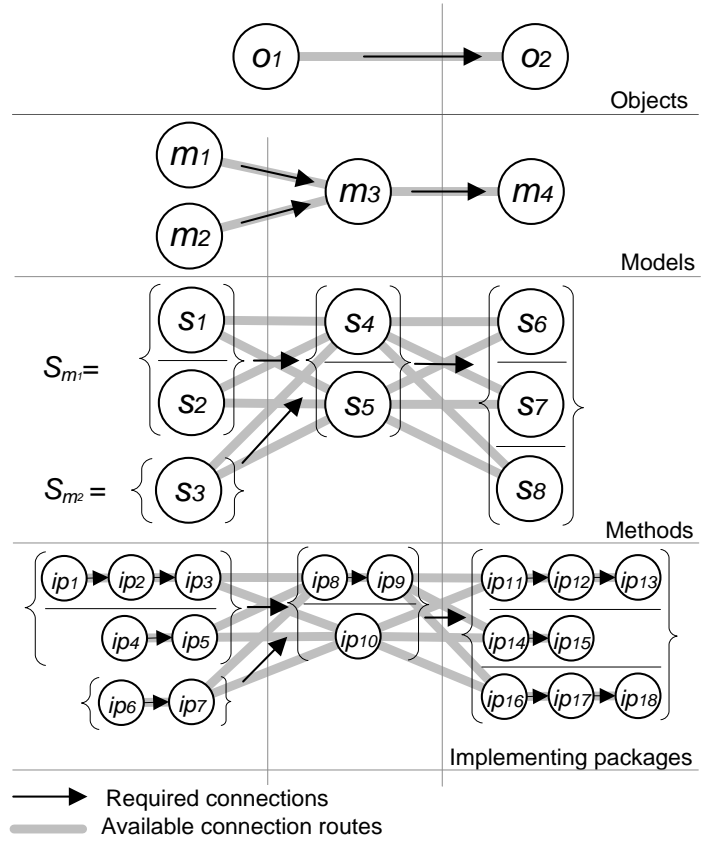

**Figure 2. VSO abstraction levels**

For initial conditions one of 12 available consequences between implementing packages $ip$ have been randomly selected. The selected consequence $(ip_4, ip_5, ip_{10}, ip_{14}, ip_{15})$ defines the following set of binary connections between the packages:

$$Conn_{IP} = \{\langle ip_4, ip_5\rangle, \langle ip_5, ip_{10}\rangle, \langle ip_{10}, ip_{14}\rangle, \langle ip_{14}, ip_{15}\rangle \quad (17)$$

These connections between implementing packages are user-defined according to semantics of package's IO parameters respectively. At the "Methods" level these packages are belong to methods $s_2, s_5, s_7$, which $IP$ sets are defined as follows:

$$IP_{s2} = \{ip_4, ip_5\}, \; IP_{s5} = \{ip_{10}\}, \; IP_{s7} = \{ip_{14}, ip_{15}\} \quad (18)$$

The associations of connected implementing packages with methods they belong to causes the following set of connections between methods:

$$Conn_S = \{\langle s_2, s_5\rangle, \langle s_5, s_7\rangle\} \quad (19)$$

At the "Models" level these methods are belong to models $m_1, m_3, m_4$, which $S$ sets are defined as follows:

$$S_{m1} = \{s_1, s_2\}, \; S_{m3} = \{s_4, s_5\}, \; S_{m4} = \{s_6, s_7, s_8\} \quad (20)$$

The association of connected methods with models they belong to brings a set of connections between models:

$$Conn_M = \{\langle m_1, m_3 \rangle, \langle m_3, m_4 \rangle\} \tag{21}$$

Objects at the top level of abstraction are defined as the following sets of simulated models:

$$M_{o1} = \{m_1, m_2, m_3\}, \ M_{o2} = \{m_4\} \tag{20}$$

The association the set of connected models with the virtual objects' they belong to gives the follows connections between objects:

$$Conn_O = \{\langle o_1, o_2 \rangle\} \tag{21}$$

So, the presented set of recursive operations prove the suggestion, that connection between elements at lower level (implementing packages is the lowest) causes the connections at upper levels. The backward statements are also true, but it will define the full amount of connections between all elements on underlying levels, which may be not correct from domain scientist's point of view. That's why elements' connection at upper levels is organized via IO parameters, obtained as a result of generalization (described before) of parameters at lower levels. Connecting objects by interconnecting their IO parameters, user connects the implemented packages at the lowest level, which is hidden from the user. A user-support regarding automatic connections between objects and models goes automatically due to user-defined semantic equivalence IO parameters, formalized during previous application design sessions.

*3. Available configurations' comparison* is an additional feature, which may be obtained thanks to semantic modeling feature of VSO concept. The feature deals with domain and problem knowledge directly. User-defined configuration hierarchical virtual object's structure defines a final workflow consequence, which will be generated as a result of application design process. The amount of final workflow variants depends geometrically depends on amount of configurations of every VSO-instance in the environment: amount of models "turnedOn" in simulation process, amount of their methods, amount of packages in selected method and etc. For example, fig. 2 demonstrates at least 12 variants of alternative workflow consequences of implementing packages, which constitute the configurations for only two virtual objects at the top level. The comparison of these consequences according to some criteria gives a valuable user-support feature for application design process. The criteria, for example, may be a total execution time or quality metrics, if the corresponding measuring models have been specified for particular packages within platform's packages definition.

## 6. Implementation & use-cases

All the logic and functionalities described above have been implemented into so-called VSO-toolbox, which consists of two GUI-applications, some amount of dynamic libraries and WCF-services, available via API. The two Silverlight-applications called VSO-Editor and VSO-environment provide a user-friendly graphical applications for the knowledge formalization, application design and workflow-code generation processes, described before. The rest VSO-components provide the reasoning functionality, which are available through the GUI-applications programmatically via WCF and http-requests. Management and storage of triples and also inference functionality are organized by Jena+Fuseki server. The triples (ontology vocabulary and facts) generation and reading are performed automatically by a generic generation/reading mechanism, which have an RDF and OWL-implementations (via OwlDotNetAPI & DotNetRdf libraries respectively). Due to reflection feature of .Net-framework the triples are generated automatically directly from C#-objects, which are visualized on screen. A modified GraphLight dynamic library is used for graph interactive visualization in VSO-environment app (nodes with different types, layout and behavior have been created). Drag&Drop operations have been implemented for nodes positioning, resizing, instantiating and connecting semantic-equal parameters from different entities (objects and models within them).

The VSO-toolbox is paired with the CLAVIRE-platform [8], which serves workflow-code interpretation, execution and monitoring processes. The formalized problem knowledge about instantiated packages required for knowledge formalization process is supplied by platform's PackageBase-service. This knowledge constitutes the lowest platform-dependent level of hierarchy (executable software packages) and becomes a base for inheritance the upper level (implementing packages). The workflow-script is the generated by toolbox automatically and transfers for execution into CLAVIRE via the platform's API. Data management functions (upload, selection from the files collection) are also based on CLAVIRE services.

A certain experiments demonstrating application of VSO toolbox for solution of different domain tasks are presented below. The first one is an experiment [3] was devoted to ship behavior simulation, depending on sea waves, which have also been simulated. As a result, some visualization has been generated using the simulated results. In paper [9] the toolbox is applied for agent-based simulations of crowd behavior, where different types of agents have been configured through virtual objects entities'. The generated configuration was used for generation of 10k agents simulating the panic in the crowd.

In case of knowledge management the toolbox have the two extra functionalities devoted to third-party knowledge integration and provenance analysis. The first one is described in paper [9] and proposes a mechanism for new VSO-images design using the existing semantic models and facts from third-party Spaql-endpoints. The idea looks perspective in case of pipeline simulation a set of complex entities. The solution of domain-specific task presented was devoted cyclones behavior simulations, which was required to tune forecast-model coefficients. About 20 cyclones found in DBpedia has been transformed into virtual object images, inheriting the generic-one. The idea of generic image is to establish some interface or schema, which should applied to all derived object-images. The created objects images have been extended with several simulation models, next have been instantiated via VSO Environment application and workflow-code have been generated. Such operations do not require user to write any program codes (parsing, processing, monitoring).

Another perspective task where VSO toolbox can be applied is a reuse existing knowledge. Paper [10] is devoted to provenance analysis, especially to extraction of consequences between steps in already executed workflows and representation these consequences onto VSO-structures (objects, models, methods, etc). This functionality brings automation into knowledge formalization process in case of automatic VSO-images creation based on already formalized domain knowledge, accumulated in provenance. During the experiments some of 169 distinct workflows have been transformed into configurable virtual objects, which are able be composed within a virtual system.

## Conclusion

In conclusion of the paper we can argue that concept and technology of Virtual Simulations Objects pairs the two types of modeling: light-weight semantic and highly-intensive simulation ones. The concept also combines the two types of knowledge: the domain and the problem one. This combination avoids the scientist to be aware about the details of particular execution platform. So VSO concept and its' implementation into VSO toolbox may be called platform-independent, because workflow consequences are generated on the fly from a composed virtual environment. The only platform dependent thing is data about bindings to the particular system's executable packages and their input/output variables. The description of VSO formal model and user-support functionalities obtained due to reasoning operations have been also contributed by this paper.

**Acknowledgement.** This paper is partially supported by Russian Scientific Foundation, grant #14-11-00823.